\documentclass[preprint2,numberedappendix]{emulateapj-rtx4}
\usepackage{graphicx,times,bm,url}
\usepackage{afterpage,lscape}
\graphicspath{{./fig/}{./png/}}

\newcommand{\EQ}{\begin{equation}}
\newcommand{\EN}{\end{equation}}
\newcommand{\EQA}{\begin{eqnarray}}
\newcommand{\ENA}{\end{eqnarray}}

\newcommand{\EEq}[1]{Equation~(\ref{#1})}
\newcommand{\Eq}[1]{Equation~(\ref{#1})}
\newcommand{\Eqs}[2]{Equations~(\ref{#1}) and~(\ref{#2})}

\newcommand{\Sec}[1]{Section~\ref{#1}}

\newcommand{\Fig}[1]{Figure~\ref{#1}}

\newcommand{\Tab}[1]{Table~\ref{#1}}

\newcommand{\meanrho}{\overline{\rho}}

\newcommand{\bec}[1]{\mbox{\boldmath $ #1$}}
{}
{}
{}
\newcommand{\meanemf}{\overline{\cal E} {}}

{}
{}
\newcommand{\meanEMF}{\overline{\mbox{\boldmath ${\cal E}$}}{}}{}
{}
{}
{}
{}
{}
\newcommand{\meanBB}{\overline{\mbox{\boldmath $B$}}{}}{}
\newcommand{\meanHH}{\overline{\mbox{\boldmath $H$}}{}}{}
{}
{}
{}
{}
{}
{}
{}
\newcommand{\meanJJ}{\overline{\mbox{\boldmath $J$}}{}}{}
\newcommand{\meanKK}{\overline{\mbox{\boldmath $K$}}{}}{}
\newcommand{\meanUU}{\overline{\bm{U}}}

{}
{}
{}

\newcommand{\meanB}{\overline{B}}

\newcommand{\meanJ}{\overline{J}}

{}

{}
{}

\newcommand{\eee}{\hat{\mbox{\boldmath $e$}} {}}

\newcommand{\xxx}{\hat{\mbox{\boldmath $x$}} {}}

\newcommand{\vv}{\mbox{\boldmath $v$} {}}

\newcommand{\xx}{\bm{x}}

\newcommand{\uu}{\mbox{\boldmath $u$} {}}
\newcommand{\UU}{\mbox{\boldmath $U$} {}}

\newcommand{\bb}{\mbox{\boldmath $b$} {}}
\newcommand{\BB}{\mbox{\boldmath $B$} {}}

\newcommand{\aaaa}{\mbox{\boldmath $a$} {}}

\newcommand{\ff}{\mbox{\boldmath $f$} {}}

\newcommand{\FF}{\mbox{\boldmath $F$} {}}

\newcommand{\grav}{\mbox{\boldmath $g$} {}}
\newcommand{\nab}{\mbox{\boldmath $\nabla$} {}}
\newcommand{\OO}{\bm{\Omega}}

\newcommand{\SSSS}{\mbox{\boldmath ${\sf S}$} {}}

\newcommand{\DD}{{\rm D} {}}

\newcommand{\dd}{{\rm d} {}}

\newcommand{\const}{{\rm const}  {}}

\def\ga{\mathrel{\mathchoice {\vcenter{\offinterlineskip\halign{\hfil
$\displaystyle##$\hfil\cr>\cr\sim\cr}}}
{\vcenter{\offinterlineskip\halign{\hfil$\textstyle##$\hfil\cr>\cr\sim\cr}}}
{\vcenter{\offinterlineskip\halign{\hfil$\scriptstyle##$\hfil\cr>\cr\sim\cr}}}
{\vcenter{\offinterlineskip\halign{\hfil$\scriptscriptstyle##$\hfil\cr>\cr\sim\cr}}}}}
\def\Ta{\mbox{\rm Ta}}
\def\Ra{\mbox{\rm Ra}}

\def\Co{\mbox{\rm Co}}

\def\Pm{\mbox{\rm Pr}_M}
\def\Rm{\mbox{\rm Re}_M}

\def\Rey{\mbox{\rm Re}}

\def\Co{\mbox{\rm Co}}

\def\cs{c_{\rm s}}
\def\kf{k_{\it f}} 

\def\urms{u_{\rm rms}}

\def\etatz{\eta_{\rm t0}}

\def\half{{\textstyle{1\over2}}}

\def\onethird{{\textstyle{1\over3}}}

\newcommand{\Myr}{\,{\rm Myr}}

\newcommand{\yapj}[3]{ #1, {ApJ,} {#2}, #3}

\newcommand{\yapjl}[3]{ #1, {ApJ,} {#2}, #3}

\newcommand{\yan}[3]{ #1, {Astron.\ Nachr.,} {#2}, #3}

\newcommand{\yana}[3]{ #1, {A\&A,} {#2}, #3}

\newcommand{\ygafd}[3]{ #1, {Geophys.\ Astrophys.\ Fluid Dyn.,} {#2}, #3}

\newcommand{\yjfm}[3]{ #1, {J.\ Fluid Mech.,} {#2}, #3}

\newcommand{\yjetp}[3]{ #1, {Sov.\ Phys.\ JETP,} {#2}, #3}

\newcommand{\ypre}[3]{ #1, {Phys.\ Rev.\ E,} {#2}, #3}

\newcommand{\yjcp}[3]{ #1, {J.\ Comput.\ Phys.,} {#2}, #3}
\newcommand{\yjour}[4]{ #1, {#2}, {#3}, #4}

\newcommand{\ybook}[3]{ #1, {#2} (#3)}

\begin{document}

\title{New scaling for the alpha effect in slowly rotating turbulence}

\author{A. Brandenburg$^{1,2}$, O. Gressel$^{1}$, P. J.\ K\"apyl\"a$^{3,1}$,
N. Kleeorin$^{4,1}$, M. J.\ Mantere$^{3}$, I. Rogachevskii$^{4,1}$
}
\affil{
$^1$Nordita, KTH Royal Institute of Technology and Stockholm
University, Roslagstullsbacken 23,
SE-10691 Stockholm, Sweden \\
$^2$Department of Astronomy, AlbaNova University Center,
Stockholm University, SE-10691 Stockholm, Sweden\\
$^3$Department of Physics, Gustaf H\"allstr\"omin katu
2a (PO Box 64), FI-00064 University of Helsinki, Finland\\
$^4$Department of Mechanical
Engineering, Ben-Gurion University of the Negev, POB 653,
Beer-Sheva 84105, Israel
}

\email{($ $Revision: 1.159 $ $)}


\begin{abstract}
Using simulations of slowly rotating stratified turbulence,
we show that the $\alpha$ effect responsible for
the generation of astrophysical magnetic fields is proportional to the
logarithmic gradient of kinetic energy density rather than
that of momentum, as was previously thought.
This result is in agreement with a new analytic theory developed
in this paper for large Reynolds numbers.
Thus, the contribution of density stratification
is less important than that of turbulent velocity.
The $\alpha$ effect and other turbulent transport coefficients are
determined by means of the test-field method.
In addition to forced turbulence, we also investigate supernova-driven
turbulence and stellar convection.
In some cases (intermediate rotation rate for forced turbulence,
convection with intermediate temperature stratification, and
supernova-driven turbulence)
we find that the contribution of density stratification
might be even less important than suggested by the analytic theory.
\end{abstract}
\keywords{magnetohydrodynamics (MHD) -- Sun:
dynamo -- turbulence}

\section{Introduction}

Turbulent dynamos occur in many astrophysical situations.
They tend to develop large-scale magnetic
structures in space and time that are generally
understood in terms of mean-field dynamo theory
\citep[e.g.][]{Mof78,Par79,KR80,ZRS83,RSS88,RH04,BS05}.
Central to this theory is the $\alpha$ effect,
which denotes a contribution to the mean electromotive force that is
given by a pseudo-scalar $\alpha$ multiplying the mean magnetic field.
Such a pseudo-scalar can be
the result of rotation, $\OO$, combined with
stratification of density and/or turbulence
intensity, $\nab\meanrho$ and/or $\nab\urms$,
respectively.
Here, $\meanrho$ is the mean gas density
and $\urms=(\overline{{\bm u}^2})^{1/2}$ is the
rms value of the turbulent velocity, ${\bm u}$.

There have been a number of analytic studies
quantifying the effects of rotating stratified
turbulence on the mean electromotive force
\citep{KR80,K91,RK93,KPR94,RKR03,KR03}. In
particular, it was found, using the quasi-linear
approach (or second-order correlation
approximation), that the diagonal components of
the $\alpha$ tensor for slow rotation rate (or
small Coriolis numbers) are given by \citep{SKR66,KR80}
\begin{eqnarray}
\alpha\approx-\ell_\alpha^2\OO\cdot\nab\ln(\meanrho \, \urms),
\label{I1}
\end{eqnarray}
where $\ell_\alpha=\tau_0\urms$ is a relevant length scale
and $\tau_0$ is the characteristic turbulent time related
to the turnover time.

In the solar convective zone the mean fluid density
$\meanrho$ changes by seven orders of magnitudes,
while the turbulent kinetic energy
($\approx \meanrho \urms^2/2$) changes by only
three orders of magnitudes and, according
to stellar mixing length theory \citep{Vit53},
$\meanrho \, \urms^3$ would be approximately constant
in the solar convective zone.
Here the fluid density $\rho$ is the sum
of the mean and fluctuating density.
This issue has become
timely, because there is a new numerical
technique that allows the different proposals to
be examined with sufficient accuracy.
The so-called test-field method \citep{Sch05,Sch07}
allows one to determine all the relevant turbulent
transport coefficients in the expression for the
mean electromotive force without
the restrictions of some of the analytic
approaches such as the quasi-linear approach,
the path-integral approach or the $\tau$ approach.
With the test-field method one solves sets of
equations for the small-scale fields resulting
from different prescribed mean fields --- the
test fields. These equations resemble
the usual induction equation, except that they
contain an additional inhomogeneous term.

This method is quite powerful because it has been shown to be rather
accurate and it gives not only the tensor coefficients of $\alpha$
effect and turbulent diffusivity, but it also allows the
scale-dependence to be determined, which means that these coefficients
are actually integral kernels that allow the effects of neighboring
points in space and time to be taken into account.  For details
regarding scale separation, see \cite{BRS08} and \cite{HB09}.  We
apply this method to numerical simulations of forced turbulence in a
stably stratified layer in the presence of rotation and a prescribed
vertical dependence of the turbulence intensity.  We also use
test-field method in simulations of turbulent convection
and supernova-driven turbulence of the interstellar medium (ISM).

The goal of the present paper is to determine the
correct scaling of the $\alpha$ effect with
mean density and rms velocity for slow rotation,
i.e., when the Coriolis number ${\rm Co}\equiv 2
\Omega \tau_0$ is much less than unity,
and large Reynolds numbers.
In addition to the parameter $\ell_\alpha$, we determine
the exponent $\sigma$ in the diagonal components
of the $\alpha$ tensor,
\begin{eqnarray}
\alpha = - \ell_\alpha^{2} \OO \cdot \nab \ln(\meanrho^\sigma\urms).
\label{I2}
\end{eqnarray}
Such an ansatz was also made by \cite{RK93}, who found
that in the high conductivity limit, $\sigma=3/2$
for slow rotation and $\sigma=1$ for rapid rotation.
However, as we will show in this paper, both numerically
(for forced turbulence and for turbulent convection with stronger
temperature stratification and overshoot layer)
as well as analytically, our results for slow rotation
and large fluid and magnetic Reynolds numbers
are consistent with $\sigma=1/2$.
In some simulations we also found $\sigma=1/3$.

\section{Theoretical predictions}

We consider the kinematic problem, i.e.,
we neglect the feedback of the magnetic field on
the turbulent fluid flow. We use a mean field
approach whereby velocity, pressure and magnetic
field are separated into mean and fluctuating parts.
Unlike in earlier derivations,
and to maintain maximum generality, we allow
the characteristic scales of the mean fluid
density, $\meanrho$, the turbulent kinetic
energy, $\meanrho \urms^2/2$,
and the variations of $\meanrho \, \urms^3$ to be different.
We also assume vanishing mean motion.
The strategy of our analytic derivation is to
determine the $\Omega$ dependencies of the second moments
for the velocity $\overline{u_{i}(t,{\bm x}) \,
u_{j}(t,{\bm x})}$ and for the cross-helicity
tensor $\overline{b_i(t,{\bm x}) \, u_{j}(t,{\bm
x})}$, where ${\bm b}$ are fluctuations of
magnetic field produced by tangling of the large-scale field.
To this end we use the equations for
fluctuations of velocity and magnetic field in
rotating turbulence, which are obtained by
subtracting equations for the mean fields from
the corresponding equations for
the actual (mean plus fluctuating) fields.

\subsection{Governing equations}

The equations for the fluctuations of velocity and
magnetic fields are given by
\begin{eqnarray}
{\partial {\bm u} \over \partial t} &=& - \cs^2
{\bec{\nabla} \rho' \over \meanrho} + 2 {\bm u} \times
{\bm \Omega} + \hat{\cal N}({\bm u}) ,
\label{A1} \\
{\partial {\bm b} \over \partial t} &=& (\meanBB
\cdot \bec{\nabla}){\bm u} - ({\bm u} \cdot
\bec{\nabla}) \meanBB - \meanBB (\bec{\nabla}
\cdot {\bm u}) + \hat{\cal N}({\bm b}),
 \label{A2} \\
{\partial\rho'\over\partial t}&=&-\nab\cdot(\meanrho {\bm u})
-\nab\cdot(\rho' {\bm u}),
\label{AA2}
\end{eqnarray}
where \Eq{A1} is written in a reference
frame rotating with constant angular velocity ${\bm \Omega}$,
and $\rho'=\rho-\meanrho$ is the fluctuating density.
We consider an isothermal equation of
state, $p=\cs^2 \rho$, so that
the density scale height is then constant.
Here, $\cs=\const$ is the sound speed,
$\meanBB$ is the mean magnetic field,
$p$ and $\rho$ are the full (mean plus fluctuating)
fluid pressure and density, respectively.
The terms $\hat{\cal N}({\bm u})$ and
$\hat{\cal N}({\bm b})$, which include nonlinear and
molecular viscous and dissipative terms, are given by
\begin{eqnarray}
\hat{\cal N}({\bm u}) &=& \overline{({\bm u} \cdot
\bec{\nabla}) {\bm u}} - ({\bm u} \cdot
\bec{\nabla}) {\bm u} + {\bm f}_{\nu}({\bm u}),
\\
\hat{\cal N}({\bm b}) &=& \bec{\nabla} \times \left({\bm u}
\times {\bm b} - \overline{{\bm u} \times {\bm
b}} - \eta \bec{\nabla} \times {\bm b} \right),
\end{eqnarray}
where $\meanrho {\bm f}_{\nu}({\bm u})$ is the molecular viscous force
and $\eta$ is the magnetic diffusion due to
the electrical conductivity of the fluid.

\subsection{The derivation procedure}

To study rotating turbulence we
perform derivations which include the
following steps:

(i) use new variables $({\bm V}, \, \meanHH)$
and perform derivations which include the
fluctuations of rescaled velocity
${\bm V} =\meanrho^\sigma \, {\bm u} $ and the mean magnetic field
$\meanHH= \meanBB / \meanrho^\sigma$;

(ii) derive equations for the second
moments of the velocity fluctuations
$\overline{V_i \, V_j}$ and the cross-helicity
tensor $\overline{b_i \, V_j}$ in ${\bm k}$ space;

(iii) apply the spectral closure, e.g.,
the spectral $\tau$ approximation
\citep{PFL76,KRR90} for large fluid and magnetic
Reynolds numbers and solve the derived
second-moment equations in ${\bm k}$ space;

(iv) return to physical space to obtain
formulae for the Reynolds stress and the
cross-helicity tensor as functions of $\Omega$.
The resulting equations allow us then to obtain
the $\Omega$ dependence of the $\alpha$ effect.
In these derivations we only take into account effects that are linear in
both ${\bm \lambda}$ and ${\bm \Omega}$, where
$\bec{\lambda}=-\bec{\nabla} \ln \meanrho$.

To exclude the pressure term from \Eq{A1}
we take twice the curl of the momentum equation
written in the new variables.
The equations which follow from \Eqs{A1}{A2}, are given by:
\begin{eqnarray}
&&{\partial \over \partial t}
\left[\bec{\nabla}^{(\lambda)}
\left(\bec{\lambda} \cdot {\bm V}\right) -
\left(\bec{\nabla}^{(\lambda)}\right)^2 {\bm V}
\right] = 2 \left({\bm \Omega} \cdot
\bec{\nabla}^{(\lambda)}\right)
\nonumber \\
&& \quad \quad \times \left(\bec{\nabla}^{(\lambda)} \times {\bm
V}\right) + 2 \left({\bm \Omega} \times
\bec{\nabla}^{(\lambda)}\right)
\left(\bec{\nabla}^{(\lambda)} \cdot {\bm
V}\right)  ,
\nonumber \\
&&\quad \quad +\hat{\cal N}({\bm V})
\label{A3} \\
&&{\partial {\bm b} \over \partial t} =
\left(\meanHH\cdot
\bec{\nabla}^{(\lambda)}\right) {\bm V} -
\left({\bm V} \cdot
\bec{\nabla}^{(\lambda)}\right) \meanHH
\nonumber \\
&& \quad \quad  + (2\sigma-1)
\left(\bec{\nabla}^{(\lambda)} \cdot {\bm
V}\right) \meanHH + \hat{\cal N}({\bm b}),
\label{A4}
\end{eqnarray}
where $\bec{\nabla}^{(\lambda)}=\bec{\nabla} +
\sigma \bec{\lambda}$.
Furthermore, $\hat{\cal N}({\bm V})=\bec{\nabla} {\bm \times}
\left(\bec{\nabla} {\bm \times} \hat{\cal N}({\bm u})\right)$
is the nonlinear term.
We consider two cases:
(i) low Mach numbers, where
the fluid velocity fluctuations ${\bm V}$ satisfy the equation
$\bec{\nabla} \cdot {\bm V} = (1-\sigma) ({\bm V}
\cdot \bec{\lambda})$ in the anelastic approximation, and
(ii) fluid flow with arbitrary Mach numbers,
where velocity fluctuations satisfy continuity equation \Eq{AA2}.
To derive \Eq{A3} we use the
identities given in Appendix~\ref{Ident}.

\subsection{Two-scale approach}

We apply the standard two-scale approach with slow and fast variables,
e.g., a correlation function,
\begin{eqnarray*}
&&\overline{V_i ({\bm x}) V_j ({\bm  y}) } = \int
\int \,d{\bm k}_1 \, d{\bm  k}_2 \,
\overline{V_i({\bm k}_1) V_j ({\bm k}_2)} \exp
\{i({\bm  k}_1 {\bm \cdot} {\bm x}
\nonumber \\
&&\quad + {\bm  k}_2 {\bm \cdot} {\bm y})\} =
\int \int\,d {\bm  k} \,d {\bm  K} \, f_{ij}({\bm
k, K}) \exp (i {\bm k} {\bm \cdot} {\bm r}+ i
{\bm K} {\bm \cdot} {\bm R})
\nonumber \\
&&\quad = \int  \,d {\bm  k} \, f_{ij}({\bm k,
R}) \exp (i {\bm k} {\bm \cdot}{\bm r}) ,
\end{eqnarray*}
\citep[see, e.g.,][]{RS75}.
Hereafter we omit the argument $t$ in the correlation
functions, $f_{ij}({\bm k, R}) = \hat L(V_i; V_j)$,
where
\begin{eqnarray*}
\hat L(a; c) = \int \overline{ a({\bm k} +
\half{\bm K}) c(-{\bm k} + \half{\bm  K}) }
\exp{(i {\bm K} {\bm \cdot} {\bm R}) } \,d {\bm
K},
\end{eqnarray*}
and we have introduced the new variables ${\bm R} =
\half({\bm x} +  {\bm y})$, ${\bm r} = {\bm x} -
{\bm y}$, $\, {\bm K} = {\bm k}_1 + {\bm k}_2$,
${\bm k} = \half({\bm k}_1 - {\bm k}_2)$. The
variables ${\bm R}$ and ${\bm K}$ correspond to
large scales, while ${\bm r}$ and ${\bm k}$
correspond to small scales. This implies
that we assume that there exists a separation of
scales, i.e., the maximum scale of turbulent
motions $\ell_0$ is much smaller than the
characteristic scale $L_B$ of inhomogeneity of
the mean magnetic field.

\subsection{Equations for the second moments}

Using \Eqs{A3}{A4} written in
${\bm k}$ space, we derive equations for the
following correlation functions: $f_{ij}({\bm
k},{\bm K}) = \overline{V_i(t,{\bm k}_1)
V_j(t,{\bm k}_2)}$ and $g_{ij}({\bm k},{\bm K}) =
\overline{b_i(t,{\bm k}_1)V_j(t,{\bm k}_2)}$. The
equations for these correlation functions are
given by
\begin{eqnarray}
{\partial f_{ij} \over \partial
t} &=& - 2 \Omega_l \Big[\varepsilon_{ipq} \Lambda_{pl}^{(1)} f_{qj}
+\varepsilon_{jpq} \Lambda_{pl}^{(2)} f_{iq} + {i \lambda_n \over k^2}
\Big(\varepsilon_{jlq} k_{q} f_{in}
\nonumber \\
&& - \varepsilon_{ilq} k_{q} f_{nj} + \varepsilon_{npq} k_{pl}
\big(k_{j} f_{iq}-k_{i} f_{qj}\big) \Big) \Big]
\nonumber \\
&& + I^f_{ij} + \hat{\cal N} f_{ij} ,
\label{A9} \\
{\partial g_{ij} \over \partial t} &=& - 2
\Omega_l \Big[\varepsilon_{jpq}
\Lambda_{pl}^{(2)} g_{iq} + {i \lambda_n \over
k^2} \Big(\varepsilon_{jlq} k_{q} g_{in}
\nonumber \\
&& + k\varepsilon_{npq} k_{plj} g_{iq} \Big) \Big]
+ i \left({\bm \Lambda}^{(1)} \cdot \meanHH\right) f_{ij}
- \overline{H}_i \lambda_n f_{nj}
\nonumber \\
&& + I^g_{ij} + \hat{\cal N} g_{ij} ,
\label{A10}
\end{eqnarray}
where $k_{ij} = k_i  k_j / k^2$, ${\bm
\Lambda}^{(1)}={\bm k}_{1} - i \sigma {\bm
\lambda}$, and similarly for ${\bm
\Lambda}^{(2)}$, $\Lambda_{mn}^{(1)}
=\Lambda_{m}^{(1)}
\Lambda_{n}^{(1)}/({\bm \Lambda}^{(1)})^2$, and
similarly for $\Lambda_{mn}^{(2)}$.
The source terms $I_{ij}^f$ and $I_{ij}^g$ in \Eqs{A9}{A10} contain
large-scale spatial derivatives of $\meanHH$,
which describe the contributions to turbulent diffusion
($I_{ij}^f$ and $I_{ij}^g$ are given by Equations~(A4) and (A6) in \cite{RK04}).
The terms $\hat{\cal N}f_{ij}$ and $\hat{\cal N}g_{ij}$ are related to the
third-order moments that are due to the nonlinear terms and are given by
\begin{eqnarray*}
\hat{\cal N} f_{ij} &=& \overline{ P_{im}({\bm
\Lambda}^{(1)}) \hat{\cal N}[V_{m}({\bm k}_1)] V_j({\bm
k}_2)}
\\
&& + \overline{V_i({\bm k}_1) P_{jm}({\bm
\Lambda}^{(2)}) \hat{\cal N}[V_{m}({\bm k}_2)]} ,
\\
\hat{\cal N} g_{ij} &=& \overline{\hat{\cal N}[b_i({\bm
k}_1)] V_j({\bm k}_2)} + \overline{b_i({\bm k}_1)
P_{jm}({\bm \Lambda}^{(2)}) \hat{\cal N}[V_{m}({\bm
k}_2)]},
\end{eqnarray*}
where $P_{ij}({\bm \Lambda}) = \delta_{ij}-\Lambda_{ij}$
and ${\bm \Lambda}={\bm k} - i \sigma {\bm \lambda}$.

\subsection{$\tau$-approach}
\label{Tau}

The equations for the second-order moments
contain higher order moments and a closure problem
arises \citep{O70,MY75,Mc90}. We apply the
spectral $\tau$ approximation or the third-order
closure procedure \citep{PFL76,KRR90,KMR96,RK04}. The
spectral $\tau$ approximation postulates that the
deviations of the third-order-moment terms,
$\hat{\cal N}f_{ij}({\bm k})$, from the
contributions to these terms afforded by the
background turbulence, $\hat{\cal
N}f_{ij}^{(0)}({\bm k})$, are expressed through
similar deviations of the second moments,
$f_{ij}({\bm k}) - f_{ij}^{(0)}({\bm k})$, i.e.,
\begin{eqnarray}
\hat{\cal N}f_{ij}({\bm k}) - \hat{\cal
N}f_{ij}^{(0)}({\bm k}) = - {f_{ij}({\bm k}) -
f_{ij}^{(0)}({\bm k}) \over \tau_r(k)} \;,
\label{A11}
\end{eqnarray}
and similarly for the tensor $\hat{\cal
N}g_{ij}$. Here the superscript $(0)$ corresponds
to the background turbulence (i.e.,
non-rotating turbulence with a zero mean magnetic
field), $\tau_r(k) $ is the characteristic
relaxation time of the statistical moments, which
can be identified with the correlation time $\tau(k)$
of the turbulent velocity field for large
Reynolds numbers. We also take into account that
$g_{ij}^{(0)}({\bm k})=0$. We apply the
$\tau$-approximation~(\ref{A11}) only to study
the deviations from the background turbulence.
The statistical properties of the background
turbulence are assumed to be known
(see below). A justification for the $\tau$
approximation in different situations has been
obtained through numerical simulations and analytical
studies \cite[see, e.g.,][]{BS05,RKB11}.

\bigskip

\subsection{Model for the background compressible turbulence}

We use the following model for the background turbulence:
\begin{eqnarray}
f_{ij}^{(0)} &\equiv& \overline{V_i^{(0)}({\bm
k}_1) \, V_j^{(0)}({\bm k}_2)} =
\Big\{\Lambda_j^{(1)} \Lambda_i^{(2)} -
\delta_{ij} \, {\bm \Lambda}^{(1)} \cdot {\bm
\Lambda}^{(2)}
\nonumber \\
&& + i \left(\lambda_i k_j -
\lambda_j k_i\right)
+ \mu_c \, \Big[k_i \, k_j + {i \over 2} \,
\big(k_i \nabla_j
\nonumber \\
&& - k_j \nabla_i\big)  \Big]\Big\} {E(k) \over
8 \pi \, k^4 (1+ \mu_c/2)} \overline{{\bm V}^2} + O(\ell_0^2 \lambda^2),
\label{A12}
\end{eqnarray}
where
\begin{eqnarray}
\mu_c = \left. \overline{ (\bec{\nabla} \cdot \, {\bm u})^2 }
\right/ \overline{(\bec{\nabla} \times {\bm u})^{2} } < 1
  \label{B40}
\end{eqnarray}
is the degree of compressibility of the turbulent velocity
field, the terms $\propto \mu_c$ take into account finite
Mach numbers compressibility effects,
$\tau(k) = 2 \tau_0 \bar \tau(k) ,$ $\,
E(k) = - d \bar \tau(k) / dk$, $\, \bar \tau(k) =
(k / k_{0})^{1-q}$, $\, 1 < q < 3$  is the
exponent of the spectrum function $(q = 5/3$ for
Kolmogorov spectrum), $k_{0} = \ell_{0}^{-1}$, $\,
\ell_{0} $ is the maximum scale of turbulent
motions, and $u_{0}$ is the characteristic
turbulent velocity at scale $\ell_{0}$.
The motion in the background turbulence is assumed to be non-helical.
For low Mach numbers ($\mu_c \ll 1)$,
\Eq{A12} satisfies to the condition:
$\bec{\nabla} \cdot {\bm V} = (1-\sigma) ({\bm V} \cdot \bec{\lambda})$.

\subsection{Contributions to the $\alpha$ effect
caused by rotation}
\label{Alpha}

Since our goal is to determine the $\alpha$ effect,
we solve \Eqs{A9}{A10}
neglecting the sources $I^f_{ij}$ and $I^g_{ij}$
with large-scale spatial derivatives of
$\meanHH$.
We subtract from \Eqs{A9}{A10} the corresponding equations
written for the background turbulence, and use the
spectral $\tau$ approximation.
We only take into account the effects which are linear in
${\bm \lambda}$ and in ${\bm \Omega}$.
We also assume
that the characteristic time of variation of the
second moments is substantially larger than the
correlation time $\tau(k)$ for all turbulence
scales. This allows us to get a stationary
solution for \Eqs{A9}{A10} for
the second-order moments.
Using this solution we determine the
contributions to the mean electromotive force
caused by rotating turbulence:
\begin{eqnarray}
\meanemf_{m} &=& \varepsilon_{mji} \, \int
\overline{b_i({\bm k}) \, V_j(-{\bm k})} \,{\rm d} {\bm k}
= \varepsilon_{mji} \, \int g_{ij}({\bm k}) \,{\rm d} {\bm k}
\nonumber \\
&=&- 2 \Omega_l \tau^2 \varepsilon_{mji} \, \int \,{\rm d} {\bm k} \, \Big[3 i
\left({\bm \Lambda}^{(1)} \cdot \meanHH\right)
\varepsilon_{jpq} \Lambda_{pl}^{(2)} f_{iq}^{(0)}
\nonumber \\
&& - \overline{H}_i \lambda_n k_{pl} \left(\varepsilon_{jpq} f_{nq}^{(0)}
+ \varepsilon_{npq} f_{qj}^{(0)}\right)
- 3{\lambda_n \over k^2}\left({\bm k} \cdot \meanHH\right)
\nonumber \\
&& \times \Big(\varepsilon_{jlq} k_{q} f_{in}^{(0)}
+\varepsilon_{npq} k_{jpl} f_{iq}^{(0)}\Big)\Big].
\label{A14}
\end{eqnarray}
After performing the integration in ${\bm k}$ space, we get:
\begin{eqnarray}
\meanemf_{i} &=& {4 \ell_0^2 \over 15}
\overline{B}_j \,  \biggl[
\left(\Omega_i \nabla_j + \Omega_j \nabla_i -
4\delta_{ij}{\bm \Omega} \cdot {\bm \nabla} \right) \ln
\overline{{\bm V}^2}
\nonumber \\
&& + (2 \sigma -1)\left(\Omega_i \lambda_j + \Omega_j \lambda_i -
4\delta_{ij}{\bm \Omega} \cdot {\bm \lambda} \right)
\nonumber \\
&& + \half \mu_c \left(\Omega_j \lambda_i -4 \Omega_i \lambda_j
+\delta_{ij}{\bm \Omega} \cdot {\bm \lambda} \right)
\biggr].
\label{A15}
\end{eqnarray}

For low Mach numbers $(\mu_c \ll 1)$ and for $\sigma=1/2$,
the $\alpha$ tensor depends only on $\nab\ln \overline{{\bm V}^2}$, i.e.,
\begin{eqnarray}
\alpha_{ij} &=& {4 \ell_0^2 \over 15}\,
\left(\Omega_i \nabla_j + \Omega_j \nabla_i -
4\delta_{ij}{\bm \Omega} \cdot {\bm \nabla} \right) \ln
\overline{{\bm V}^2},
\label{A17}
\end{eqnarray}
where we used $\meanemf_{i}=a_{ij} \overline{B}_j$, and
$\alpha_{ij}\equiv\half(a_{ij}+a_{ji})$
is the symmetric part of $a_{ij}$.
Furthermore, the pumping velocity, $\gamma_i\equiv\half
\varepsilon_{inm} a_{mn}$ is independent of rotation
for small Coriolis numbers, because the rotational contribution
to the pumping velocity vanishes.
In this case the pumping velocity
is determined only by the inhomogeneity
of turbulent magnetic diffusivity.

For arbitrary Mach numbers and when
\begin{eqnarray}
\sigma={1 \over 2} + {\mu_c \over 16},
\label{A25}
\end{eqnarray}
the diagonal part of the $\alpha$ effect depends
only on $\nab\ln \overline{{\bm V}^2}$, i.e.,
\begin{eqnarray}
\alpha &=& - {16 \ell_0^2 \over 15}\,
{\bm \Omega} \cdot {\bm \nabla} \ln
\overline{{\bm V}^2} .
\label{A26}
\end{eqnarray}
The latter equation can be rewritten in the
following form:
\begin{eqnarray}
\alpha &=& - {32 \ell_0^2 \over 15}\,
{\bm \Omega} \cdot {\bm \nabla} \ln
\left(\meanrho^\sigma \, \urms\right) ,
\label{AA26}
\end{eqnarray}
where $\sigma$ is determined by \Eq{A25}.
In the next section we will determine the exponent
$\sigma$ from numerical simulations.

\section{Numerical simulations}

We now aim to explore the universality of the derived scaling by
comparison with results from very different astrophysical
environments. To do this, we performed three quite different types of
simulations of:

(i) artificially forced turbulence of a rotating stratified gas, where
the density scale height is constant and the turbulence is driven by
plane wave forcing with a given wavenumber;

(ii) supernova-driven interstellar turbulence in a
vertically-stratified local Cartesian model employing the shearing
sheet approximation;

(iii) turbulent convection with and without overshoot layers.

The advantage of the first approach is that it allows to impose a
well-defined vertical gradient of turbulent intensity, i.e., of the
rms velocity of the turbulence such that $\nab\ln\urms$ is approximately
constant over a certain $z$ interval, excluding the region near the
vertical boundaries. Naturally, the physically more complex scenarios
(ii) and (iii) are less-well controlled but allow to demonstrate the
existence of the $\alpha$~effect scaling in applications of direct
interest to the astrophysical community. We use different variants of
the test-field method to measure the $\alpha$ effect (and other
turbulent transport coefficients).

\subsection{Basic equations}

In the following we consider simulations where the turbulence is either driven
by a forcing with a function $\ff$, in which case we assume an isothermal gas
with constant sound speed $\cs$, or it is driven through heating and
cooling either by supernovae in the interstellar medium or through
convection with heating from below.
In all cases, we solve the equations for the velocity, $\UU$,
and the density $\rho$
in the reference frame rotating with constant
angular velocity ${\bm \Omega}$ and linear shear rate $S$:
\begin{eqnarray}
\rho{\DD\UU\over\DD t}&=&-\nab p+\rho(\ff_\Omega+\ff+\grav)
+\nab\cdot(2\nu\rho\SSSS),\\
{\partial\rho\over\partial t}&=&-\nab\cdot(\rho\UU),
\end{eqnarray}
where $\ff_\Omega=(2\Omega U_y,(S-2\Omega)U_x,0)$ is a combined Coriolis and
tidal acceleration for $\Omega$ pointing in the $z$ direction and
a linear shear flow $\UU_S=(0,Sx,0)$, $\nu$ is the kinematic viscosity,
$\ff$ is a forcing function, $\grav$ is gravity, and
${\sf S}_{ij} =\half(\nabla_j U_i+\nabla_i U_j)
-\onethird\delta_{ij}\nab\cdot\UU$ is the
traceless rate-of-strain tensor, not to be confused with the shear rate $S$,
and $\DD/\DD t=\partial/\partial t+(\UU+\UU_S)\cdot\nab$ is the advective
derivative with respect to the total (including shearing) velocity.
In the isothermal case, the pressure is given by $p=\rho\cs^2$,
while in all other cases we also solve an energy equation,
for example in terms of the specific entropy $s=c_v\ln p-c_p\ln\rho$,
where $c_p$ and $c_v$ are respectively the specific heats at constant
pressure and constant volume, their ratio $\gamma=c_p/c_v$ is chosen
to be 5/3, and $s$ obeys
\begin{equation}
\rho T{\DD s\over\DD t}=-\nab\FF_{\rm rad}+\rho\Gamma-\rho^2\tilde\Lambda
+2\nu\rho\SSSS^2,
\end{equation}
where the temperature $T$ obeys $(c_p-c_v)T=p/\rho$,
$\FF_{\rm rad}$ is the radiative flux, $\Gamma$ is a heating function,
and $\tilde\Lambda$ is a cooling function.
In the isothermal case, the entropy equation is not used, and
the forcing function $\ff$ consists of random,
white-in-time, plane, non-polarized waves with a
certain average wavenumber, $\kf$.

The simulations are performed with the {\sc Pencil Code}
(\url{http://pencil-code.googlecode.com})
which uses sixth-order explicit finite differences in space and a
third-order accurate time stepping method \citep{BD02}.
The simulations of supernova-driven turbulence in the ISM have been
performed using the {\sc Nirvana}-III code \citep{2004JCoPh.196..393Z}
with explicit viscosity and resistivity.

\subsection{The test-field method}

We apply the kinematic test-field method
\citep[see, e.g.,][]{Sch05,Sch07,BRS08} to determine
all relevant turbulent transport coefficients in the general relation
\EQ
\meanemf_i=\alpha_{ij}\meanB_j+\eta_{ijk}\meanB_{j,k},
\EN
where $\meanB_{j,k}=\nabla_k \meanB_{j}$ is the magnetic gradient tensor.
The test-field method works with a set of test fields $\meanBB^T$,
where the superscript $T$ stands for the different test fields.
The corresponding mean electromotive forces $\meanEMF^T$
are calculated from $\meanEMF^T=\overline{\uu \times \bb^T}$,
where $\bb^T=\nab\times\aaaa^T$ with
\EQ
{\partial\aaaa^T\over\partial t}=\meanUU\times\bb^T+\uu\times\meanBB^T+
(\uu\times\bb^T)'+\eta\nabla^2\aaaa^T \, .
\label{eq051}
\EN
Here, $\meanUU$ and $\uu$ taken from the solutions of the momentum equation.
In the case with shear, we replace $\partial\aaaa^T/\partial t$ by
$\partial\aaaa^T/\partial t+\UU_S\cdot\nab\aaaa^T+Sa^T_y\xxx$.
On the top and bottom boundaries we assume perfect conductors, and
for the $x$ and $y$ directions periodic boundary conditions.
These small-scale fields are then used to determine the electromotive
force $\meanEMF^T$ corresponding to the test field $\meanBB^T$.
The number and form of the test fields used depends on the problem at hand.

We either use planar ($xy$) averages, which depend only on $z$ and $t$
(hereafter referred to as test-field method~I),
or, alternatively, we assume that the mean field varies also in the $x$
and $y$ directions, but that the turbulence is homogeneous in those two
directions, and that the $z$ direction constitutes a preferred direction
of the turbulence (test-field method~II).
In the former case, I, only the $x$ and $y$ components of $\meanEMF$ are important
for dynamo action, and the magnetic gradient tensor has only two non-vanishing
components which can be expressed in terms of the components of the mean current
density alone.
We put $\meanJJ=\nab\times\meanBB$ such that $\meanJJ/\mu_0$, with $\mu_0$
being the magnetic permeability, is the mean current density.
Thus, we have
\begin{equation}
\meanemf_i=\alpha_{ij}\meanB_j - \eta_{ij}\meanJ_j,
\end{equation}
with $i$ and $j$ being either 1 or 2.

Alternatively, in case II, the mean electromotive force is assumed to be characterized by
only one preferred direction which we describe by the unit vector $\eee$.
Then, $\meanEMF$ can be represented in the form
\EQA
\meanEMF&=&
\alpha_\perp\meanBB
+(\alpha_\parallel-\alpha_\perp)(\eee\cdot\meanBB)\eee
+\gamma\eee\times\meanBB
\nonumber \\
&&-\eta_\perp\meanJJ
-(\eta_\parallel-\eta_\perp)(\eee\cdot\meanJJ)\eee
-\delta\eee\times\meanJJ
\label{eq005}\\
&&-\kappa_\perp\meanKK
-(\kappa_\parallel-\kappa_\perp)(\eee\cdot\meanKK)\eee
-\mu\eee\times\meanKK
\nonumber
\ENA
with nine coefficients $\alpha_\perp$, $\alpha_\parallel$, $\ldots$, $\mu$.
Like $\meanJJ = \nab \times \meanBB$, also $\meanKK$
is determined by the gradient tensor $\nab \meanBB$.
While $\meanJJ$ is given by its antisymmetric part,
$\meanKK$ is a vector defined by $\meanKK=\eee \cdot (\nab\meanBB)^\mathrm{S}$
with $(\nab\meanBB)^\mathrm{S}$ being the symmetric part of $\nab \meanBB$.
For details of this method, referred to below as test-field method II,
see \cite{BRK12}.
\EEq{I2} is expected to apply to $\alpha_\perp$, while
$\alpha_\parallel$ can, in certain cases,
have opposite sign \citep{BNPST90,Fer92,RK93}.
This is why we associate in the following $\alpha$ in this equation
with the $\alpha_\perp$ defined in \Eq{eq005}.

Errors are estimated by dividing the time series into three equally
long parts and computing time averages for each of them. The largest
departure from the time average computed over the entire time series
represents an estimate of the error.

\subsection{Simulations of forced turbulence}

We begin by studying forced turbulence.
We consider a domain of size $L_x\times L_y\times L_z$ in Cartesian
coordinates $(x,y,z)$, with periodic boundary
conditions in the $x$ and $y$ directions and
stress-free, perfectly conducting boundaries at
top and bottom, $z=\pm L_z/2$.
The gravitational acceleration, $\grav=(0,0,-g)$, is
chosen such that the density scale height $H_\rho=\cs^2/g$
is small compared with the vertical extent of the domain, i.e., $L_z$.
The smallest wavenumber that fits into the cubic
domain of size $L^3$ is $k_1=2\pi/L$, so the density
contrast between bottom and top is
$\exp(2\pi)\approx535$ and the mean density varies
like $\meanrho=\meanrho_0\exp(-z/H_\rho)$, where $\meanrho_0$
is a constant.
In all cases, we use a scale separation ratio $\kf/k_1=5$,
a fluid Reynolds number $\Rey\equiv\urms/\nu\kf$ between 60 and 100,
a magnetic Prandtl number $\Pm=\nu/\eta$ of unity.
We use a numerical resolution of $128^3$ mesh points
for all forced turbulence runs.

We perform simulations for different values of the rms velocity gradient,
$\lambda^{(u)}\equiv\dd\ln\urms/\dd z$, but fixed logarithmic density gradient,
$\dd\ln\meanrho/\dd z\equiv-\lambda$.
Thus, we have
\EQ
\alpha_\perp=\ell_\alpha^2\Omega\lambda\left(\sigma-\lambda^{(u)}/\lambda\right),
\EN
i.e., $\sigma$ can be obtained conveniently as the value
of $\lambda^{(u)}/\lambda$ for which $\alpha$ vanishes.
By arranging the turbulence such that $\lambda^{(u)}$ and $\lambda$
are approximately independent of $z$, the value of $\alpha$ is also
approximately constant.
In that case, however, all the other turbulent transport coefficients
are $z$-dependent.
However, by normalizing $\gamma$ by $\urms/6$ and the other coefficients
by $\etatz(z)=\urms(z)/3\kf$, we obtain non-dimensional quantities that
are approximately independent of $z$.
We denote the corresponding non-dimensional quantities by a tilde
and quote in the following their average values over an interval
$z_1\leq z\leq z_2$, in which these ratios are approximately constant.

\begin{figure}[ht!]\begin{center}
\includegraphics[width=\columnwidth]{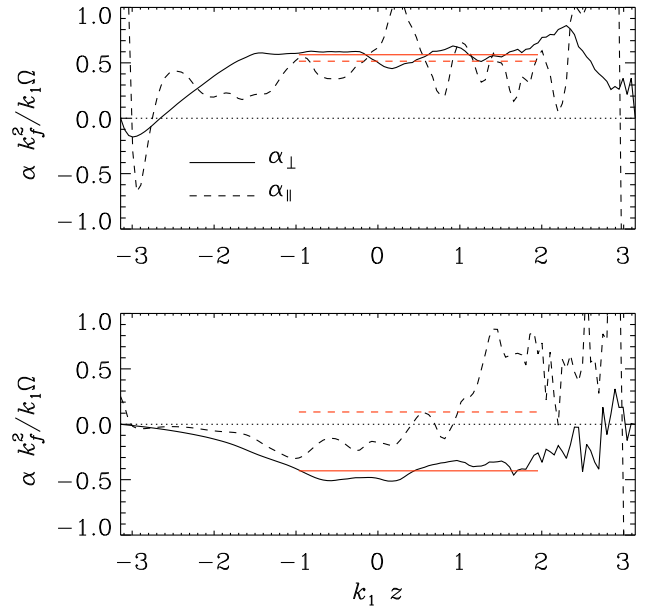}
\end{center}\caption[]{
Comparison of $\alpha_\perp$ (solid) and $\alpha_\parallel$ (dashed) for
the runs with $\lambda^{(u)}=0.33$ (upper panel) and 0.61 (lower panel).
}\label{palpcomp}\end{figure}

In \Fig{palpcomp}, we plot the normalized profiles of
$\tilde\alpha_\perp\equiv\alpha_\perp\kf^2/k_1\Omega$ and
$\tilde\alpha_\parallel\equiv\alpha_\parallel\kf^2/k_1\Omega$ as functions of $z$.
Note that within the range $z_1\leq z\leq z_2$ with $k_1z_1=-1$ and $k_1z_2=2$,
both functions are approximately constant.
For $\lambda^{(u)}\ga0.5$, they are of opposite sign; see the lower panel
of \Fig{palpcomp} and \Tab{Tresults}.
As discussed above, this behavior has been seen and interpreted
in earlier calculations \citep{BNPST90,Fer92,RK93}.

\begin{figure}[ht!]\begin{center}
\includegraphics[width=\columnwidth]{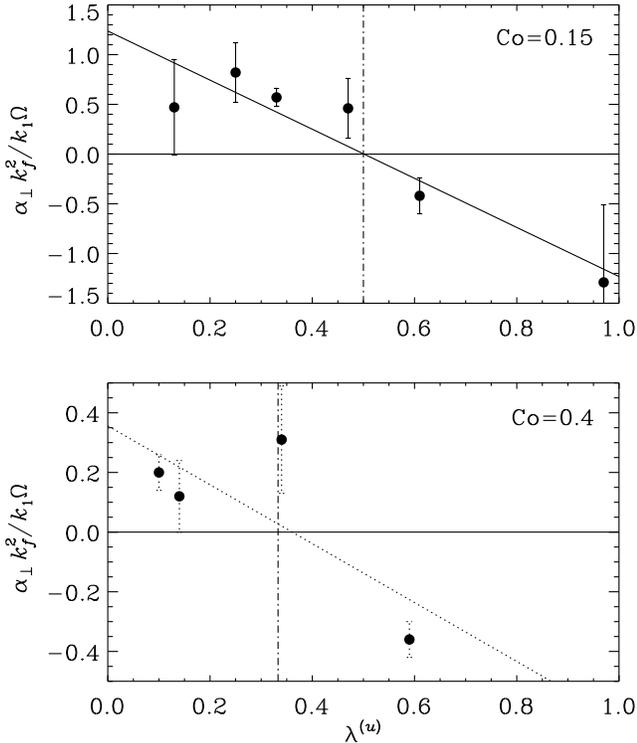}
\end{center}\caption[]{
Dependence of the normalized mean values of $\alpha_\perp$
on the value of $\lambda^{(u)}$
for $\Co=0.15$ (upper panel) and 0.4 (lower panel),
giving respectively $\sigma\approx1/2$ and $\approx1/3$
as the zeros in each graph.
}\label{presults}
\end{figure}

In \Fig{presults} we plot the dependence of the normalized mean values
of $\alpha_\perp$ on the value of $\lambda^{(u)}$ for two values of $\Co$.
The value of $\sigma$ can then be read off as the zero of that graph.
We find $\sigma\approx1/2$ for low values of $\Co$, and a somewhat
smaller value ($\sigma\approx1/3$) for larger values of $\Co$.

\begin{figure}[ht!]\begin{center}
\includegraphics[width=\columnwidth]{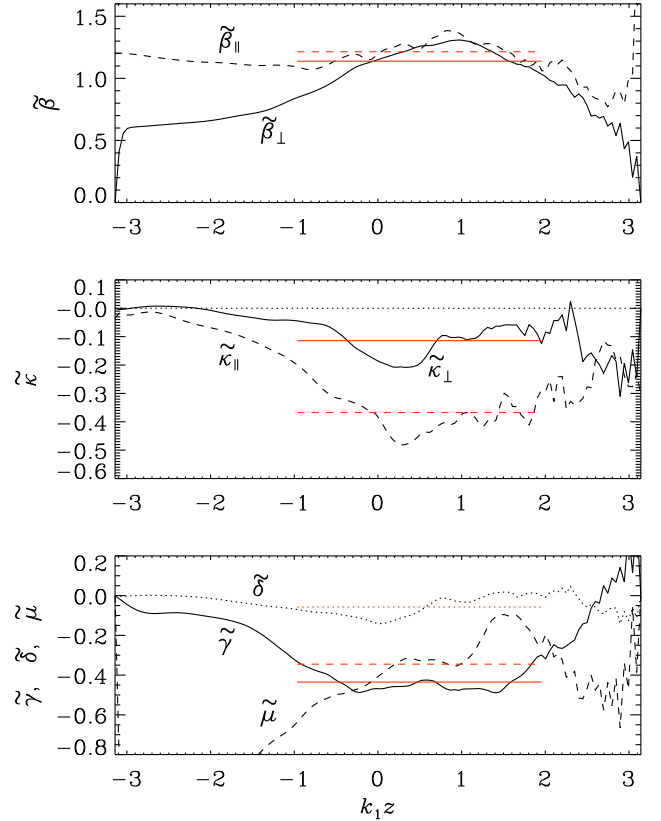}
\end{center}\caption[]{
Transport coefficients for the run with $\lambda^{(u)}=0.61$.
}\label{prestSexp10c}\end{figure}

\begin{figure}[ht!]\begin{center}
\includegraphics[width=\columnwidth]{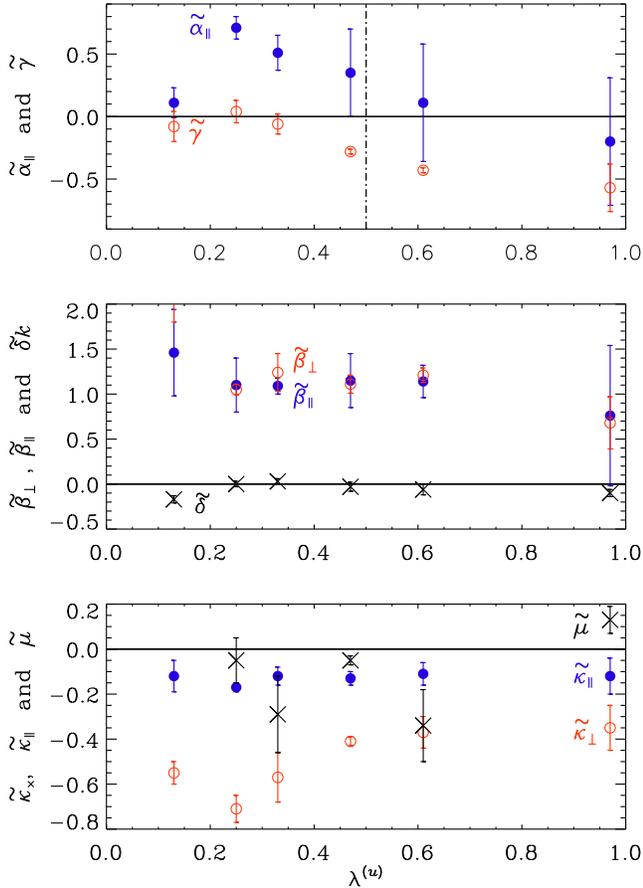}
\end{center}\caption[]{
Results for the other transport coefficients for $\Co=0.15$.
}\label{presults_other}
\end{figure}


\begin{table*}[htb]\caption{
Basic parameters and turbulent transport coefficients for the
forced turbulence simulations.
}\vspace{12pt}\centerline{\begin{tabular}{rrrrrrrrrrrrrrrrr}
\hline
\hline
$\Rm$ & $\Co$ & $\lambda^{(u)}$
& $\alpha_{\perp}\kf^2/\Omega k_1$ & $\alpha_{\parallel}\kf^2/\Omega k_1$ & $\gamma/\etatz(z) k_1$
& $\beta_{\perp}/\etatz(z)$ & $\beta_{\parallel}/\etatz(z)$ & $\delta/\etatz(z)$
& $\kappa_{\perp}/\etatz(z)$ & $\kappa_{\parallel}/\etatz(z)$ & $\mu/\etatz(z)$ \\
\hline
 70&0.14&0.13&$ 0.47\pm0.16$&$ 0.1\pm0.1$&$-0.1\pm0.1$&$ 1.5\pm0.2$&$ 2.1\pm0.3$&$-0.2\pm0.0$&$-0.1\pm0.1$&$-0.6\pm0.1$&$-1.8\pm0.4$\\
 56&0.17&0.25&$ 0.82\pm0.10$&$ 0.7\pm0.1$&$ 0.0\pm0.1$&$ 1.1\pm0.1$&$ 1.0\pm0.1$&$ 0.0\pm0.0$&$-0.2\pm0.0$&$-0.7\pm0.1$&$-0.1\pm0.1$\\
 62&0.16&0.33&$ 0.57\pm0.03$&$ 0.5\pm0.1$&$-0.1\pm0.1$&$ 1.1\pm0.2$&$ 1.2\pm0.2$&$ 0.0\pm0.0$&$-0.1\pm0.0$&$-0.6\pm0.1$&$-0.3\pm0.2$\\
 47&0.20&0.46&$ 0.32\pm0.26$&$ 0.2\pm0.3$&$-0.2\pm0.1$&$ 1.1\pm0.2$&$ 1.0\pm0.2$&$-0.0\pm0.0$&$-0.1\pm0.0$&$-0.4\pm0.0$&$-0.1\pm0.0$\\
 65&0.15&0.61&$-0.42\pm0.06$&$ 0.1\pm0.5$&$-0.4\pm0.0$&$ 1.1\pm0.1$&$ 1.2\pm0.1$&$-0.1\pm0.1$&$-0.1\pm0.1$&$-0.4\pm0.1$&$-0.3\pm0.2$\\
 59&0.16&0.97&$-1.29\pm0.26$&$-0.2\pm0.5$&$-0.6\pm0.2$&$ 0.8\pm0.3$&$ 0.7\pm0.3$&$-0.1\pm0.0$&$-0.1\pm0.1$&$-0.3\pm0.1$&$ 0.1\pm0.1$\\
\hline
108&0.36&0.10&$ 0.20\pm0.02$&$-0.1\pm0.1$&$ 0.0\pm0.0$&$ 0.9\pm0.1$&$ 1.6\pm0.2$&$ 0.1\pm0.0$&$ 0.1\pm0.0$&$-0.5\pm0.1$&$-1.6\pm0.2$\\
 93&0.41&0.14&$ 0.12\pm0.04$&$-0.1\pm0.0$&$-0.0\pm0.0$&$ 1.0\pm0.1$&$ 2.2\pm0.4$&$ 0.0\pm0.0$&$ 0.1\pm0.0$&$-0.3\pm0.1$&$-1.7\pm0.4$\\
 75&0.51&0.35&$ 0.30\pm0.05$&$ 0.3\pm0.0$&$-0.2\pm0.1$&$ 1.0\pm0.1$&$ 0.8\pm0.1$&$ 0.0\pm0.0$&$-0.3\pm0.0$&$-1.1\pm0.0$&$ 0.2\pm0.0$\\
 74&0.52&0.59&$-0.36\pm0.02$&$ 0.2\pm0.1$&$-0.7\pm0.1$&$ 1.2\pm0.0$&$ 0.8\pm0.0$&$-0.3\pm0.1$&$-0.2\pm0.1$&$-0.9\pm0.0$&$ 0.3\pm0.0$\\
\hline
\hline
\label{Tresults}\end{tabular}}
\end{table*}


In \Fig{prestSexp10c} we show the $z$-dependence of the remaining
7 normalized coefficients.
They are all approximately independent of $z$ within the
same range $z_1\leq z\leq z_2$ as before.
In \Fig{presults_other} we show the scaling of these turbulent
transport coefficients (including now also $\tilde\alpha_\parallel$)
with $\lambda^{(u)}$.
The results for $\alpha_\parallel$ and $\gamma$ suggest a
dependence proportional to the gradient of
$\meanrho^{\sigma_i}\urms$
with $\sigma_{\alpha_\parallel}$ between 0 and 0.3 for $\alpha_\parallel$ and
$\sigma_\gamma$ between 0.7 and 1 for $\gamma$.
On the other hand, all the other coefficients seem to be independent
of $\lambda$ and we find
$\tilde\beta_\parallel\approx\tilde\beta_\perp\approx1.1$,
$\tilde\delta\approx0$,
$\tilde\kappa_\parallel\approx-0.1$, $\tilde\kappa_\perp\approx-0.4$,
and $\tilde\mu\approx-0.05$.
These results are quantitatively and qualitatively in agreement
with those of \cite{BRK12}.
The fact that $\tilde\delta$ turned out to be essentially zero was
addressed earlier \citep{BRK12}, where it was found that significant
values are only found for scale separation ratios around unity,
i.e., when the scale of the mean field is comparable to that of
the turbulent eddies.

\subsection{The case of supernova-driven ISM turbulence}
\label{ISM}

\begin{figure}
\begin{center}
  \includegraphics[width=0.95\columnwidth]{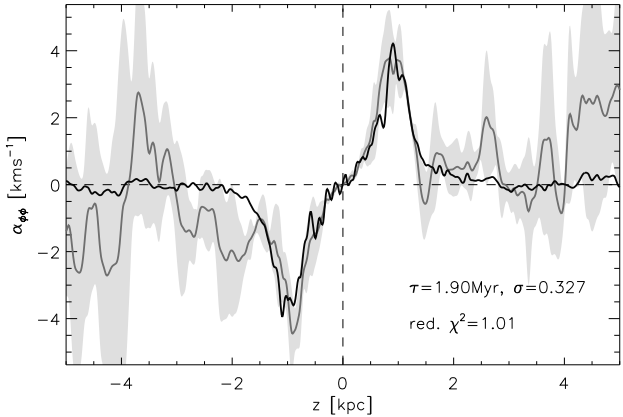}
  \end{center}
  \caption{Time-averaged vertical profile of $\alpha_{yy}$ from
    ISM turbulence
    \citep[see][]{2011IAUS..274..348G}, obtained with the
    test-field method (gray line). The best-fit model (black line)
    with $\sigma=0.327$ is obtained by method of least squares,
    weighted with the standard deviation (shaded area) in $\alpha_{yy}$.
    \label{fig:ism}}
\end{figure}

We now turn to simulations of supernova-driven ISM turbulence
\citep[see][for a detailed description of the
  model]{2008AN....329..619G,2008A&A...486L..35G},
similar to those of \cite{Korpi_etal99} and
\cite{2012arXiv1206.6784G}, but extending to larger box
sizes, thus allowing for better scale separation. In
these simulations, expansion
waves are driven via localized injection of thermal energy,
$\Gamma_{\rm SN}(\xx,t)$. Additionally, optically thin radiative cooling
with a realistic cooling function $\tilde \Lambda(T)$ and heating
$\rho\Gamma(z)$ lead to a segregation of the system into multiple ISM phases.

Here we solve the visco-resistive compressible MHD equations
(supplemented by a total energy equation), using the {\sc Nirvana}-III
code \citep{2004JCoPh.196..393Z}; for the full set of equations, we
refer the reader to Equations~(2.1) in \citet{2010arXiv1001.5187G}.
For the present run, we chose a resolution of $128\times128\times512$
mesh points, and apply a value of $\Pm=2.5$. The fluid Reynolds
number, defined as $\Rey\equiv \urms\,\ell_0/(2\pi \nu)$, varies
within the domain and takes values $\Rey\simeq 70$--$165$, which is
somewhat larger than for the forced turbulence case,
where $\ell_0=2\pi/\kf$.
The Coriolis number ${\rm Co}\equiv 2 \Omega \tau_0$ is here $\simeq 0.24$.

Despite the aforementioned differences, the basic
properties of the turbulence producing an $\alpha$ effect are in fact
quite similar: rotation of the system together with stratification in
the mean density and turbulence amplitude. Notably, the ISM simulations are
strongly compressible with peak Mach numbers of up to ten,
corresponding to a typical value of $\mu_c\simeq 1.9$, and with peak
values up to five. We note that the ISM simulations also include shear
with $S=-\Omega$, which may affect the production of vorticity.

\begin{figure}
  \center\includegraphics[width=0.95\columnwidth]{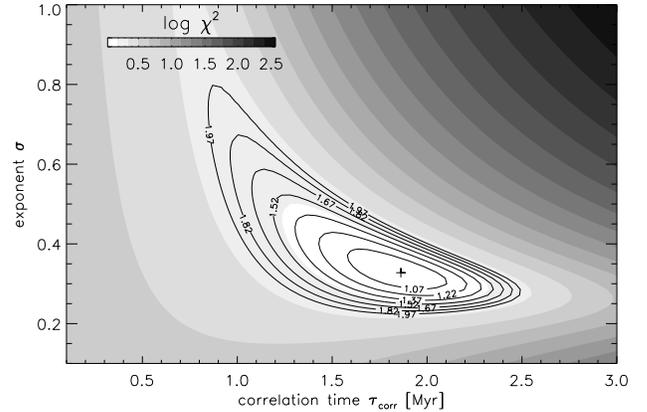}
  \caption{Likelihood map based on the reduced-$\chi^2$ error
    estimate, with the best-fit parameter set indicated by the cross.
    \label{fig:fit_soca_map}}
\end{figure}

To obtain an estimate for $\sigma$ from the time-series of a single
simulation run, we apply a method distinct from the one described
above: We treat the $\alpha_{yy}(z)$ profile (inferred with
test-field method~I) as the \emph{data} to be modeled and obtain error
estimates by means of the standard deviation within four equal
sub-intervals in time (see gray line and shaded areas in \Fig{fig:ism}).
We then compute time averages of the profiles for $\nab\ln\meanrho$ and
$\nab\ln\urms$ (here without error estimates), from
which we compute a \emph{model} prediction for $\alpha_{yy}(z)$ based
on the expression
\begin{equation}
  \alpha_{yy} = - \tau_\alpha^2 u_{\rm rms}^2\,
  \OO\cdot\nab\ln \left( \meanrho^\sigma u_{\rm rms}  \right),
  \label{eq:fit}
\end{equation}
where we have assumed that $\ell_\alpha$ can be replaced by
$\tau_\alpha u_{\rm rms}(z)$,
with $\urms(z)$ being the $z$-dependent rms velocity and $\tau_\alpha$
(assumed independent of $z$) is a characteristic timescale related to
the $\alpha$~effect. We apply a least-square optimization
allowing $\tau_\alpha$ and $\sigma$ as free parameters.
The best-fit model according to \Eq{eq:fit} is plotted as a black line
in \Fig{fig:ism}, along with the stated values for $\tau_\alpha$
and $\sigma$, and matches well the data within the error bars.

Because we cannot {\it a priori} assure that $\tau_\alpha$
is uniform in space, and because this might affect the precise
determination of $\sigma$, we perform an additional test. We do this
by independently estimating $\ell_0(z)$ from the two-point velocity
correlation function, computed in horizontal slabs around a given
galactic height $z$, and time averaged over multiple snapshots.
By comparing the obtained $\ell_0(z)$ with $u_{\rm rms}(z)$, we find that
our data are broadly compatible with a uniform $\tau_0$ of about $1.2\Myr$.
To corroborate the fit, we compute a likelihood map in the
parameter space spanned by $\tau_\alpha$ and $\sigma$ and find that
the best-fit parameter set is, in fact, located at the global minimum
of the reduced-$\chi^2$ map; see \Fig{fig:fit_soca_map}.
The best-fit value of $\tau_\alpha$ is around $1.9\Myr$, which is
indeed compatible with the correlation time $\tau_0$.

To conclude this section, we remark that we here
find a somewhat smaller exponent,
$\sigma\simeq 1/3$, which suggests that this case deviates from the
theoretical prediction.
It has however a similar exponent as in
our stratified forced turbulence simulations with larger values of
$\Co$. Note that with the determined value for $\mu_c$, \Eq{A25}
predicts a value of $\sigma\simeq 0.62$, which is a factor two larger
than obtained from the fit.
The reason for such discrepancy between the theoretical predictions and
the simulations of supernova-driven ISM turbulence might be caused by the fact
that the theory is developed for simplified conditions which
are different from these simulations.

\subsection{Convection-driven turbulence}

\begin{figure}[ht!]\begin{center}
\includegraphics[width=\columnwidth]{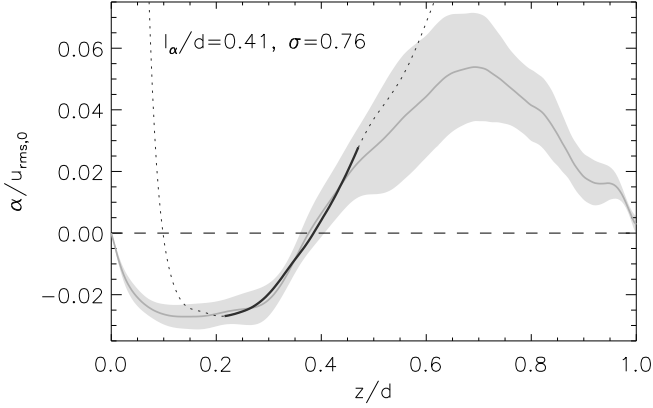}
\end{center}\caption[]{
  Dependence of $\alpha$ on $z$ for convective turbulence without
  overshoot layers (gray line, with shaded areas indicating
  fluctuations) compared with \Eq{I2} (black line / dotted) applying
  $\sigma=0.76$ and $\ell_\alpha=0.41\,d$, as obtained from a
  least-square fit within the highlighted interval in
  $z/d$.}\label{palpeta}
\end{figure}

\begin{figure}[ht!]\begin{center}
\includegraphics[width=\columnwidth]{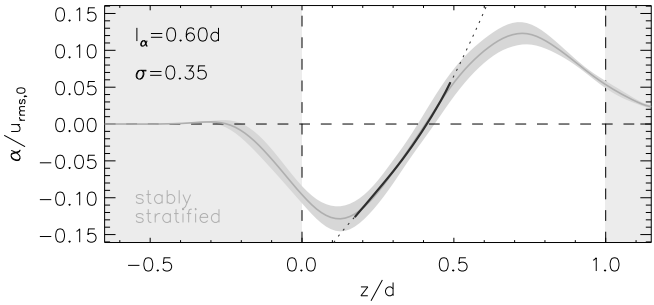}
\end{center}\caption[]{
Same as \Fig{palpeta}, but the case with overshoot layers and weak
temperature stratification, comparing $\alpha$ (gray line, shaded
areas) measured via the test-field method with \Eq{I2} (black line / dotted)
yielding $\sigma=0.35$ and $\ell_\alpha=0.6\,d$ as best-fit values.
}\label{palpeta_tf128a1}
\end{figure}

\begin{figure}[ht!]\begin{center}
\includegraphics[width=\columnwidth]{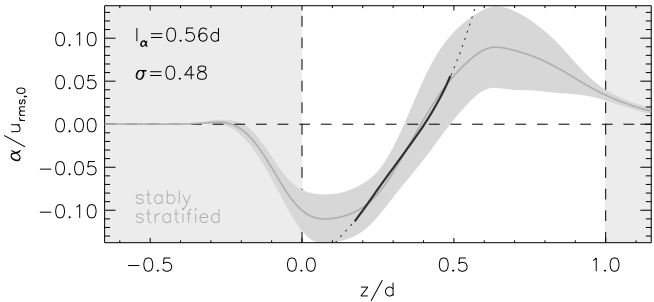}
\end{center}\caption[]{
Same as \Fig{palpeta}, but the case with overshoot layers
and stronger temperature stratification, comparing
$\alpha$ (gray line, shaded areas) with \Eq{I2}
using $\sigma=0.48$ and $\ell_\alpha=0.56\,d$ (black line / dotted).
}\label{palpeta_tf128a2}
\end{figure}

\begin{table}[b!]\caption{Summary of convection runs.
}\vspace{12pt}\centerline{\begin{tabular}{cccrrcl}
\hline
\hline
Run & $\Ra$ & $\Ta$ & $\Delta \meanrho_{\rm total}$ & $\Delta \meanrho_{\rm CZ}$ & overshoot & Res. \\
\hline
A & $1.3\cdot10^6$ & $3.6\cdot10^5$ &  64 &  64 & $-$ & $256^3$ \\
B & $6.1\cdot10^5$ & $6.4\cdot10^4$ &  37 &   7 & + & $128^3$ \\
C & $6.1\cdot10^5$ & $6.4\cdot10^4$ & 290 & 296 & + & $128^3$ \\
\hline
\hline
\label{convectionruns}\end{tabular}}
\end{table}

Many astrophysical bodies have turbulent convection zones.
Again, rotation and stratification induce helicity into the flow
and therefore drive an $\alpha$ effect.
In stellar mixing length theory \citep{Vit53}, one assumes that the
temperature fluctuation, $T'$, is proportional to $\urms^2$, so
the convective flux $\overline{(\rho\uu)'c_p T'}$ is well approximated by
$\meanrho\urms^3$, which is also confirmed by simulations \citep{BCNS05}.
In the steady state, the total energy flux is constant in space,
so if most of the flux is carried by convection,
then $\meanrho\urms^3=\const$
and thus its vertical gradient vanishes.
If the scaling of \Sec{ISM} were applicable also to this case, i.e., if
$\alpha\propto\dd\ln\meanrho\urms^3/\dd z$, then $\alpha$ would vanish.
To investigate this somewhat worrisome possibility, we now consider
Now we consider
a simulation of turbulent convection in a stratified layer,
heated from below by a constant energy flux $\FF_{\rm rad}$
at the bottom, where we adopt the diffusion approximation for an
optically thick gas with $\FF_{\rm rad}=-K\nab T$ and radiative
conductivity $K$.
We either consider a constant value of $\chi=K/(\rho c_p)$ with an
enhanced turbulent heat conductivity $\chi_{\rm t}$ near the surface as
in spherical simulations of \cite{KMB11}, or, alternatively,
a piecewise constant profile $K(z)$, such that ${\bm g}\cdot\nab s$ is positive
in the middle of the domain, which corresponds to convective instability.
The latter setup is described in detail in \cite{KKB09}.
The hydrostatic equilibrium value of ${\bm g}\cdot\nab s$ is proportional
to the Rayleigh number, $\Ra$, which is here around $10^6$;
see \cite{KKB08} for the definition.
The rotational influence is here measured in terms of the Taylor
number $\Ta=(2\Omega d^2/\nu)^2$, which is $3.6 \cdot 10^5$ for Run A.
Summary of our convection simulations is given in \Tab{convectionruns}.
Run~A without overshoot layers and $6.4\cdot10^3$ for Runs~B and C.
Here $d$ is the thickness of the unstable layer.
The density contrast, $\meanrho_{\max}/\meanrho_{\min}$,
in the run without overshoot layers is 64.
The vertical boundary conditions are stress-free and we use $\Pm=0.5$
and $\Rm=52$ in Run~A, and $\Pm=1$ and $\Rm=13$ in Runs~B and C.
In \Fig{palpeta} we give the results for Run~A without overshoot layers.
The Coriolis number varies with height and is about 0.2 in the middle
of the layer.
The rotation axis is anti-parallel to the direction of gravity,
corresponding thus to a location at the north pole.
The system is therefore isotropic in the $xy$ plane and we consequently
quote the mean between the two horizontal components, i.e.,
$\alpha=\half(\alpha_{xx}+\alpha_{yy})$, using test-field method~I,
which corresponds to $\alpha_\perp$ of test-field method~II.
Note that in this simulation, $\alpha(z)$ shows a sinusoidal profile,
suggestive of either weak stratification or effects of boundaries.
Given that density stratification is not small
(the density contrast is 64), it is plausible
that the effects of boundaries are here responsible for the extended
regime with negative $\alpha$.

Our simulation shows that the best fit is obtained for $\sigma\approx0.75$.
A possible reason for this unexpected behavior might be poorer scale
separation in convection simulations compared with forced turbulence
simulations.
The other possibility is related to the absence of convective overshoot
layers discussed above.
This idea is partly confirmed by comparing with simulations that include
convective overshoot layers.
Now the best fit value for $\sigma$ is found to be about 1/3 when the temperature
stratification is weak (\Fig{palpeta_tf128a1}) and about 1/2 when the temperature
stratification is strong (\Fig{palpeta_tf128a2}).

\begin{table}[t!]\caption{Summary of results for $\sigma$.
}\vspace{12pt}\centerline{\begin{tabular}{llrc}
\hline
\hline
 & $\Co$ & $\meanrho_{\max}/\meanrho_{\min}$ & $\sigma$  \\
\hline
forced turbulence & 0.15 & 535 & 1/2 \\
                  & 0.40 & 535 & 1/3 \\[2pt]\hline
supernova-driven ISM & 0.24  & 1000& 1/3 \\[2pt]\hline
convective turbulence (CT) & 0.2  &  64 & 3/4 \\
CT with overshoot   & 0.2  &  37 & 1/3 \\
                      & 0.2  & 290 & 1/2 \\[2pt]\hline
analytic theory       &      &     & 1/2 \\
\hline
\hline
\label{summaryall}\end{tabular}}
\end{table}

\section{Conclusions}

While the present investigations confirm the old result that the
$\alpha$ effect in mean-field dynamo theory emerges as the combined action
of rotation and stratification of either density or of turbulent intensity,
they also now point toward a revision of the standard formula for
$\alpha$.
The old formula by \cite{SKR66} predicted that the effect of stratification
can be subsumed into a dependence on
the gradient of $\meanrho\urms$.
This formula was then generalized by \cite{RK93} to a dependence on
$\meanrho^\sigma\urms$, where $\sigma=3/2$
in the high conductivity limit for slow rotation,
and $\sigma=1$ for faster rotation.
In contrast, our new results now clearly favor a value of $\sigma$
below unity.
The idealized case of artificially forced turbulence can most directly
be compared to our analytic derivation, since it agrees in all the
made assumptions.
The obtained value of
$\sigma=1/2$ agrees very well with the theoretical expectation. A
similar exponent is found for the case of turbulent convection with
higher temperature stratification, but the results seem to depend
sensitively on model parameters (see~\Tab{summaryall}).
Here more detailed studies will be
required.
Moreover, the result $\sigma=1/2$ arises naturally from analytical
considerations for large fluid and magnetic Reynolds numbers and slow rotation
as the only tenable choice,
but those considerations have not yet been performed for
the cases of intermediate and rapid rotation.

Forced turbulence
simulations show a trend toward smaller values of $\sigma$ around 1/3
for faster rotation and also in cases of supernova-driven turbulence.
Turbulent convection with overshoot also gives 1/3
in one case of moderate temperature
stratification with overshoot, while simulations without overshoot point
toward values somewhat larger values around 3/4.
However, in none of the cases we have found that the $\alpha$ effect
diminishes to zero as a result of a trend toward constant convective
flux for which $\meanrho\urms^3$ is approximately constant.
In spite of the considerable scatter of the values of $\sigma$ found
from various simulations, it is worth emphasizing that in all cases
$\sigma$ is well below unity.
On theoretical grounds, the value 1/2 is to be expected.
Except for the forced turbulence simulations that also yield 1/2 for
slow rotation, all other cases are too complex to expect agreement
with our theory that ignores, for example, inhomogeneities of the
density scale height and finite scale separation.

\acknowledgments

We thank Karl-Heinz R\"adler for detailed comments on our manuscript.
We acknowledge the NORDITA dynamo programs of
2009 and 2011 for providing a stimulating
scientific atmosphere. Computing resources
provided by the Swedish National Allocations
Committee at the Center for Parallel Computers at
the Royal Institute of Technology in Stockholm,
the High Performance Computing Center North
in Ume{\aa}, and CSC -- IT Center for Science in
Espoo, Finland.
This work was supported in part by
the European Research Council under the AstroDyn
Research Project No.\ 227952
(AB), by COST Action MP0806, by the European
Research Council under the Atmospheric Research
Project No.\ 227915, by a grant from the
Government of the Russian Federation under
contract No.\ 11.G34.31.0048 (NK,IR), Academy of
Finland grants No. 136189, 140970 (PJK)
and 218159, 141017 (MJM), and the University of
Helsinki `Active Suns' research project.
Part of this work used the {\sc Nirvana} code version 3.3,
developed by Udo Ziegler at the Leibniz-Institut
f{\"u}r Astrophysik Potsdam (AIP).

\appendix

\section{Identities used for the derivation of \Eq{A3}}
\label{Ident}

To derive \Eq{A3} we use the following
identities:
\begin{eqnarray}
&&\meanrho^\sigma \, \left[\bec{\nabla} {\bm \times}
\left(\bec{\nabla} {\bm \times} {\bm u}\right)
\right]_i = \left[\nabla_i^{(\lambda)}
\nabla_j^{(\lambda)} -\delta_{ij} \,
\big(\bec{\nabla}^{(\lambda)}\big)^2 \right] V_j,
\label{A5}\\
&&\meanrho^\sigma \, \big[\bec{\nabla} {\bm \times}
[\bec{\nabla} {\bm \times} ({\bm u} {\bm \times}
{\bm \Omega})] \big]_i =
\left[\nabla_i^{(\lambda)} \nabla_j^{(\lambda)}
-\delta_{ij} \,
\big(\bec{\nabla}^{(\lambda)}\big)^2 \right]
\nonumber \\
&& \quad \quad \times ({\bm V} {\bm \times} {\bm \Omega})_j,
\label{A6}\\
&&\left[\nabla_i^{(\lambda)} \nabla_j^{(\lambda)}
-\delta_{ij} \,
\big(\bec{\nabla}^{(\lambda)}\big)^2 \right]
({\bm V} {\bm \times} {\bm \Omega})_j
=\big({\bm\Omega}  {\bm \times}
\bec{\nabla}^{(\lambda)}\big)_i (\bec{\lambda}
{\bm \cdot} {\bm V})
\nonumber \\
&& \quad \quad + \big({\bm \Omega} {\bm \cdot}
\bec{\nabla}^{(\lambda)}\big)
\big(\bec{\nabla}^{(\lambda)}  {\bm \times} {\bm
V}\big)_i . \label{A7}
\end{eqnarray}
Equation~(\ref{A7}) is obtained by multiplying
the identity
\begin{eqnarray}
\varepsilon_{ijm} \Omega_m  + \Omega_l
\left(\varepsilon_{jml} \Lambda_{im} -
\varepsilon_{iml} \Lambda_{jm}\right)=
\varepsilon_{ijm} \Lambda_{ml} \Omega_l,
\label{A8}
\end{eqnarray}
by $\Lambda^2 V_j$, where $\varepsilon_{ijk}$ is the
fully antisymmetric Levi-Civita tensor,
$\Lambda_{mn} =\Lambda_{m} \Lambda_{n}/\Lambda^2$, and the
identity~(\ref{A8}) is valid for arbitrary
vectors ${\bm \Omega}$ and ${\bm \Lambda}$.


\end{document}